\documentclass{article}

\usepackage[preprint]{neurips_2024}

\usepackage{amsmath,amsfonts,bm}

\def\eqref#1{equation~\ref{#1}}
\def\1{\bm{1}}

\DeclareMathAlphabet{\mathsfit}{\encodingdefault}{\sfdefault}{m}{sl}
\SetMathAlphabet{\mathsfit}{bold}{\encodingdefault}{\sfdefault}{bx}{n}

\usepackage{tabularray}

\usepackage{amssymb}% http://ctan.org/pkg/amssymb
\usepackage{pifont}% http://ctan.org/pkg/pifont
\usepackage{rotating}
\usepackage{makecell}
\usepackage{graphicx}
\usepackage{chngpage}
\usepackage{float}
\usepackage{hyperref}
\usepackage{subfigure}
\usepackage{tabularx}
\usepackage{dutchcal}
\usepackage{multirow}
\usepackage{amssymb}
\usepackage{pifont}
\usepackage{booktabs} % For professional looking tables
\usepackage{array}
\usepackage{longtable}
\usepackage{tabularx}
\usepackage{geometry} % Optional, for adjusting page margins if needed
\usepackage{multirow} % For multirow feature

\newcolumntype{L}[1]{>{\raggedright\let\newline\\\arraybackslash\hspace{0pt}}m{#1}}
\newcolumntype{C}{>{\centering\arraybackslash}p{0.18\linewidth}} % Adjust width as necessary

\usepackage[utf8]{inputenc} % allow utf-8 input
\usepackage[T1]{fontenc}    % use 8-bit T1 fonts
\usepackage{hyperref}       % hyperlinks
\usepackage{url}            % simple URL typesetting
\usepackage{booktabs}       % professional-quality tables
\usepackage{amsfonts}       % blackboard math symbols
\usepackage{nicefrac}       % compact symbols for 1/2, etc.
\usepackage{microtype}      % microtypography
\usepackage{xcolor}         % colors

\title{Detecting Popular Social Events through Limited Observation with Deep Survival Analysis}

\author{%
  Maryam Ramezani\\
  Sharif University of Technology\\
  \texttt{maryam.ramezani@sharif.edu}
  \And
  Hossein Goli\\
  Sharif University of Technology\\
  \texttt{hossein.goli@sharif.edu}
  \And
  AmirMohammad Izadi\\
  Sharif University of Technology\\
  \texttt{amirmohammad.izadi01@sharif.edu}
  \And
  Hamid R. Rabiee\\
  Sharif University of Technology\\
  \texttt{rabiee@sharif.edu}
}

\begin{document}

\maketitle

\begin{abstract}
Users increasing activity across various social networks made it the most widely used platform for exchanging and propagating information among individuals. To spread information within a network, a user initially shared information on a social network, and then other users in direct contact with him might have shared that information. Information expanded throughout the network by repeatedly following this process. A set of information that became popular and was repeatedly shared by different individuals was called popular trends. Identifying and analyzing these trends led to valuable insights into the dynamics of information dissemination within a network. However, more importantly, proactive approaches emerged. In other words, by observing the dissemination pattern of a piece of information in the early stages of expansion, it became possible to determine whether this cascade would become highly popular in the future. This research aimed to predict and detect popular trends in social networks by observing limited early-stage data and using a deep survival analysis-based method. This model could play a significant role in improving recommendation systems, predicting the reach of digital content, and assisting in optimal decision-making in digital marketing. Ultimately, the proposed method was tested on various real-world anonymized datasets from Twitter, Weibo, and Digg.\end{abstract}

\section{Introduction}\label{sec:Introduction}
The increasing acceptance of users of all kinds of social networks has made it the most used platform to exchange and propagate information between people. Social networking activities, such as posting, tweeting, retweeting, sharing, quoting, liking, commenting, engaging, and voting on a piece of content, including news, videos, images, or memes on the various social platforms, result in ``social events''. To propagate information over a network, a user initially publishes information on a social network, then other users in his direct contact may be infected and share that information. Information spreads across the network by repeating this process. 
The process by which information spreads over the network is known as a ``cascade" and is a set of events occurring across the network. A certain amount of information is so popular that it is shared repeatedly by different individuals over time and is referred to as a ``viral cascade''.

Identifying and analyzing viral social events can provide valuable insight into the dynamics of information propagation over a network so that stakeholders can quickly adopt appropriate strategies. However, predictive approaches are more important than identification. In other words, by observing the propagation pattern of a piece of information at the earliest stage of the spread, it is possible to determine whether or not the information will go viral. 

The detection of viral events in a social network is a powerful tool for the management of information and for making informed decisions.
In social cascades, the detection of viral events is useful for a broad range of applications, some of which include the following.
\begin{itemize}
    \item 
    Influence prediction: Various information strategies can be evaluated by identifying trends about whether current news or content on social networks will go viral to achieve the most significant impact on user behavior.
    \item 
    Supply and demand analysis: To properly manage the supply of goods and services, it is essential to predict the amount of advertising done among social network users. If the demand for the product exceeds the threshold, the warehouse will have sufficient stock to meet it.
    \item 
    Decision-making in marketing: By recognizing the virality of various methods, companies can measure the effectiveness of their advertising or marketing campaigns and make better marketing decisions.
    \item 
    Promoting adequate information: Identifying the extent of virality early in a news broadcast can enhance the transmission of more impactful information to society. This proactive approach fosters increased social engagement in problem-solving and opinion-sharing.
    % \item 
    % Early detection of rumors: Rumors are information with speculative and unverified characteristics. A rumor may be completely false, partially true, or completely accurate in terms of accuracy. When a rumor spreads among users, one of the most critical questions is whether this rumor is a viral event or whether it is merely a non-viral rumor. If a rumor does not infect many users, it is unnecessary to invest time and money to prevent its spread. However, if we know that many users will be exposed to this rumor, it is necessary to manage the crisis and take steps to eliminate its effects or prevent its spread. In cases where this rumor is false information or fake news, the issue becomes more sensitive. We must adopt mitigation approaches if we understand that it will likely become viral.

\end{itemize}

Viral events detection extends beyond social networks to any complex network that can spread signals of information over its communication links, including: 

\begin{itemize}
    % \item 
    % Viral disease prevalence: When a virus spreads through a population, an estimate of virality can be made based on the number of infections at different times. Thus, it can be determined whether there is a need to quarantine or restrict drug production shortly. 
    \item 
    Traffic congestion: Transport networks include communication links between spatial nodes. Observations of commuting at different times can be used to predict whether traffic and crowding will exceed a defined controllable limit in future hours. Control policies for traffic management must be applied before congestion occurs if the crowd is expected to exceed the predefined limit.
\end{itemize}

There are many challenges in detecting viral cascades: 
\begin{enumerate}
    \item 
 Limited data: We can determine the virality of a cascade of events by counting the number of infections if the cascade lifetime is completely observed. However, the main challenge arises when only part of the data is available. This makes monitoring the limited early stages of the cascade life cycle necessary to detect the viral cascades. Studies have shown that viral and non-viral cascades may have the same pattern in the early stages, making it difficult to distinguish between various types of cascades \cite{EPOC2022}. 
    \item 
    Dependent data: When a user is exposed to information, it comes from his following users. So, the number of infected users at a time depends on the number of infected users at the previous time. In addition, the cascade pattern has a “quick rise-and-fall” feature at different times \cite{matsubara2012rise}. How to consider these features in learning the pattern of cascades is a challenge \cite{EPOC2022, wang2015burst}.
    \item
    Limited features:
    Some aspects of social networking can be difficult or time-consuming to access. Gathering the content of information, users' profiles, following and follower links and the content of other users' profiles is a large amount of data that is not fully accessible due to platform policies. As a result of the inherent characteristics of social networks, the path of information propagation is not apparent. This only shows the source of the cascade, i.e., the publisher's account, not who forwarded it in the last step. In addition, the hidden links between the graphs of websites are another example of impossible data.
\end{enumerate}

The rest of the paper is organized as follows. 
The definitions and preliminary concepts are briefly discussed in Section 2. 
Section 3 describes the formal definition of the problem addressed in this paper. Section 4 briefly overviews the entire framework and introduces the two main stages of our framework. Section 5 presents the experimental design and the results of the evaluation.
Finally, we conclude our work in the last section.

\section{Definitions and Preliminary Concepts}\label{sec:DefinitionsAndPreliminaryConcepts}
% \begin{definition}
\textbf{Viral Events}: When user $u$ disseminates information $i$ over the network at a certain time $t$, a social Event is generated, represented by $(u,t)$. The sequence of social events associated with a particular piece of information $i$ over a network is called a Social Event Cascade. The cascade $c_{i}$ about information ${i}$ with $n$ number of user infections is defined as $c_{i}=\{(u_0, t_0), (u_1, t_1), \dots, (u_{n-1}, t_{n-1}), (u_n, t_n)\}$ where $t_0 \le t_1 \le \dots \le t_{n}$ and $u_0$ is the initiator (source) of the cascade. Several of these initiator events are more likely to receive greater attention and reach a wider audience. The cascade of events, in this case, is known as the ``viral'' phenomenon. Formally, a Viral Social Event is an event $(u_0,t_0)$ that initiates a cascade $c_{i}$ in which the amount of infected users ($n$) exceeds a certain threshold of $\zeta$.

\textbf{Survival Analysis}: Survival analysis is a method of modeling events over time using statistical methods. Let $T$ be a continuous nonnegative random variable that indicates the time of a certain event. 
The probability of the event occurring exactly at time $t$, which is called the probability density function (PDF), is:
\begin{equation}\label{eq:f(t)}
    f(t) = \lim_{dt \rightarrow 0} \frac{P(t \le T < t+dt)}{dt}
\end{equation}
The Cumulative Distribution Function (CDF) gives the probability of the event time occurring earlier than or equal to the time $t$:
\begin{equation}\label{eq:F(t)}
    F(t) = P(T \le t) = \int f(t)d(t)
\end{equation}
A hazard function measures the probability that an event will occur at time $t$ given it has not already occurred. 
\begin{equation}\label{eq:hazard}
    h(t) = \lim_{dt \rightarrow 0} \frac{P(t \le T<t+dt|T\geq t)}{dt} 
\end{equation}
% \end{definition}

% \begin{definition}
\textbf{Covariate}: The term covariate in survival analysis refers to a variable associated with the time it takes for an event to occur. They are used to investigate the impact of different factors on survival time or hazard rate. User profiles, the number of followers a user has, the content of the information, user reactions, time of activities, network structure, or any other relevant variables that could influence the time to an event are covariate examples.
% \end{definition}

% \begin{definition}
\textbf{Survival Function}:     
The survival function gives the probability that the occurrence time of an event is longer than some specified time $t$:
\begin{equation}\label{eq:S(t)}
    S(t) = P(T \geq t) = 1 - F(t)
\end{equation}
With conditional probability, we can write:
% \begin{equation}
% \frac{P(t \le T<t+dt|T\geq t)}{dt}=\frac{P(t \le T<t+dt,t \le T)}{P(T\geq t)dt} 
% =\frac{P(t \le T<t+dt)}{S(t)dt}
% \label{eq:conditionalprob}
% \end{equation}
\begin{align}\label{eq:conditionalprob}
\begin{aligned}
\frac{P(t \le T<t+dt|T\geq t)}{dt} &= \frac{P(t \le T<t+dt,t \le T)}{P(T\geq t)dt} \\
&= \frac{P(t \le T<t+dt)}{S(t)dt}
\end{aligned}
\end{align}
Taking limits from both sides of the Equation \ref{eq:conditionalprob}, with using Equations \ref{eq:f(t)} and \ref{eq:hazard}:
\begin{equation} \label{eq:survival_hazard_pdf}
    h(t) = \frac{f(t)}{S(t)}
\end{equation}
Derivation of both sides of Equation \ref{eq:S(t)}, then using Equation \ref{eq:F(t)}: 
\begin{equation} \label{eq:derivation_S(t)}
    \frac{dS(t)}{dt}=\frac{-dF(t)}{dt}=-f(t) 
\end{equation}
Equations \ref{eq:survival_hazard_pdf} and \ref{eq:derivation_S(t)} leads to the core formula of survival function:
\begin{equation} \label{eq:survival_hazard_core}
    h(t) = \frac{f(t)}{S(t)} = \frac{-dS(t)}{dt} \times \frac{1}{S(t)}
\end{equation}
By integrating from both sides of \eqref{eq:survival_hazard_core}:
\begin{equation} \label{eq:survival_hazard_integral}
    \int -\frac{dS(t)}{S(t)} = \int h(t)dt
\end{equation}
In conclusion, we can derive the Survival function depending only on the hazard function at previous times. The final equation in continuous space is:
\begin{equation} \label{eq:survival_hazard_continuous}
      S(t) = \exp{\left(-\int_{0}^{t}h(x)dx\right)}
\end{equation}
Subsequently, if the problem is defined in discrete space:
\begin{equation} \label{eq:survival_hazard_discrete}
    S(t) = \exp{\left(-\sum_{k=1}^{t}h_k\right)}
\end{equation}
% \end{definition}
% \begin{definition}
\textbf{Censored Data}: Survival analysis may be incomplete if the time to event is unknown, but we know it will not occur within the observation window. These missing data are referred to as right-censored data.
% \end{definition}

\begin{table}[t]
\centering % This centers the table
\caption{Notation}\label{tbl1}
\begin{tabular*}{0.9\linewidth}{@{\extracolsep{\fill}} p{0.15\linewidth} p{0.75\linewidth} }
\toprule
\textbf{Notation} & \textbf{Description}\\
\midrule
D        & Number of training cascades \\ 
T        & Time of Viral of a random variable \\ 
$f(t)$   & Probability density function \\ 
$F(t)$   & Cumulative density function \\
$S(t)$   & Survival Function                       \\ 
$h(t)$ & Hazard rate function          \\ 
$\zeta$ & Viral threshold          \\ 
$C_i$ & $i^{th}$ cascade       \\ 
$t_i$ & $i^{th}$ reshare time in a Cascade     \\ 
$a_{i,t}$ & Label indicating whether cascade $i$ has Viral until time $t$\\ 
$L$ & Maximum Length of Cascade     \\ 
$\lambda$ & Weibull Distribution lambda parameter     \\
$k$ &  Weibull Distribution K parameter     \\ 
\bottomrule
\end{tabular*}
\end{table}

\section{Problem Definition}\label{sec:ProblemDefinition}
\begin{figure*}[t]
    \centering
    \includegraphics[width=1\textwidth]{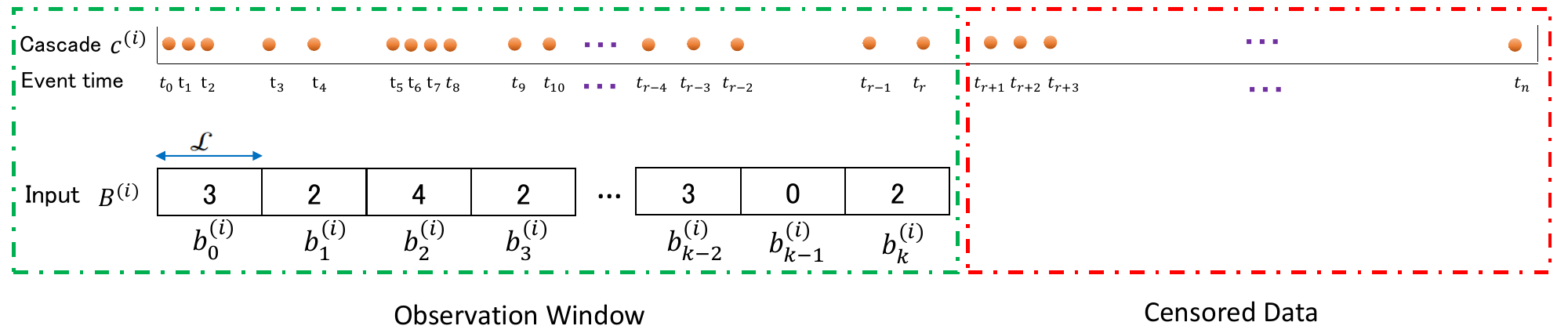}
    \caption{The cascades have censored data. We Divide the observed part of cascade $c^{(i)}$ into multiple bins $b^{(i)}_{j}$s with length $\mathcal{L}$ for generating the input of the model. Each bin counts the number of events.}
    \label{fig:input}
\end{figure*}

Given a censored cascade $c^{(i)}$, which observation window with length $r$, our task is to predict whether it will be a viral cascade in the future life of its propagation $v^{(i)}=1$ or not $v^{(i)}=0$. In other words, using the first $r$ events of a cascade, will the number of events from the beginning to the end of cascade age $n$ be greater or less than the defined threshold $\zeta$.
As mentioned, a cascade is a sequence of social events. We rely only on the most present data, the time of the event, to be robust to missing data problems for collecting different social information. Consequently, this problem does not utilize other data, such as event content, user profiles, or propagation paths. As presented in Figure \protect \ref{fig:input}, to transform the input to a proper discrete form named $B^{(i)}$, we will divide each $c^{(i)}$ into multiple bins with length $\mathcal{L}$. Then the value of bin $b^{(i)}_{j}$ will contain the number of events in that bin. The goal of this paper is a binary classification for learning function $\rho(B^{(i)}) \rightarrow v^{(i)}$.

\newpage

\section{Proposed Method}\label{sec:ProposedMethod}
Since cascades represent a set of social events, they are easily modeled using survival analysis, which gives the probability that the occurrence time of becoming viral will be longer than some specified time.  
We can classify viral and non-viral cascades according to the experience of previous studies \cite{adamic2017} by estimating the survival function from uncensored data. Our primary concern is to learn the survival function based on the three cases of (1) using minimal data features, (2) learning from a minimal amount of uncensored data, and (3) compensating for this lack of information by employing data dependence. Therefore, the bins of $B^{(i)}$ are the covariates of this survival analysis. 

Mathematically speaking, the proposed method learns a function $\rho$ composed of two functions $\rho = \delta o \gamma$. 
As illustrated in Figure \protect \ref{fig:overview}: First, $\gamma(B^{(i)})$ fits the survival function, capturing long-term dependencies in cascades that are significant to predict an appropriate survival function $S(t)$. The $\delta(S(t))$ discriminator is then based on this inferred survival function and predicts the label of the sequence $v^{(i)}$.
In this section, we describe the proposed method ``VEDSA'' for the detection of viral events with deep survival analysis using learning $\rho$ composed of two models:

\begin{figure*}[ht]
\centering
\includegraphics[scale=0.41]{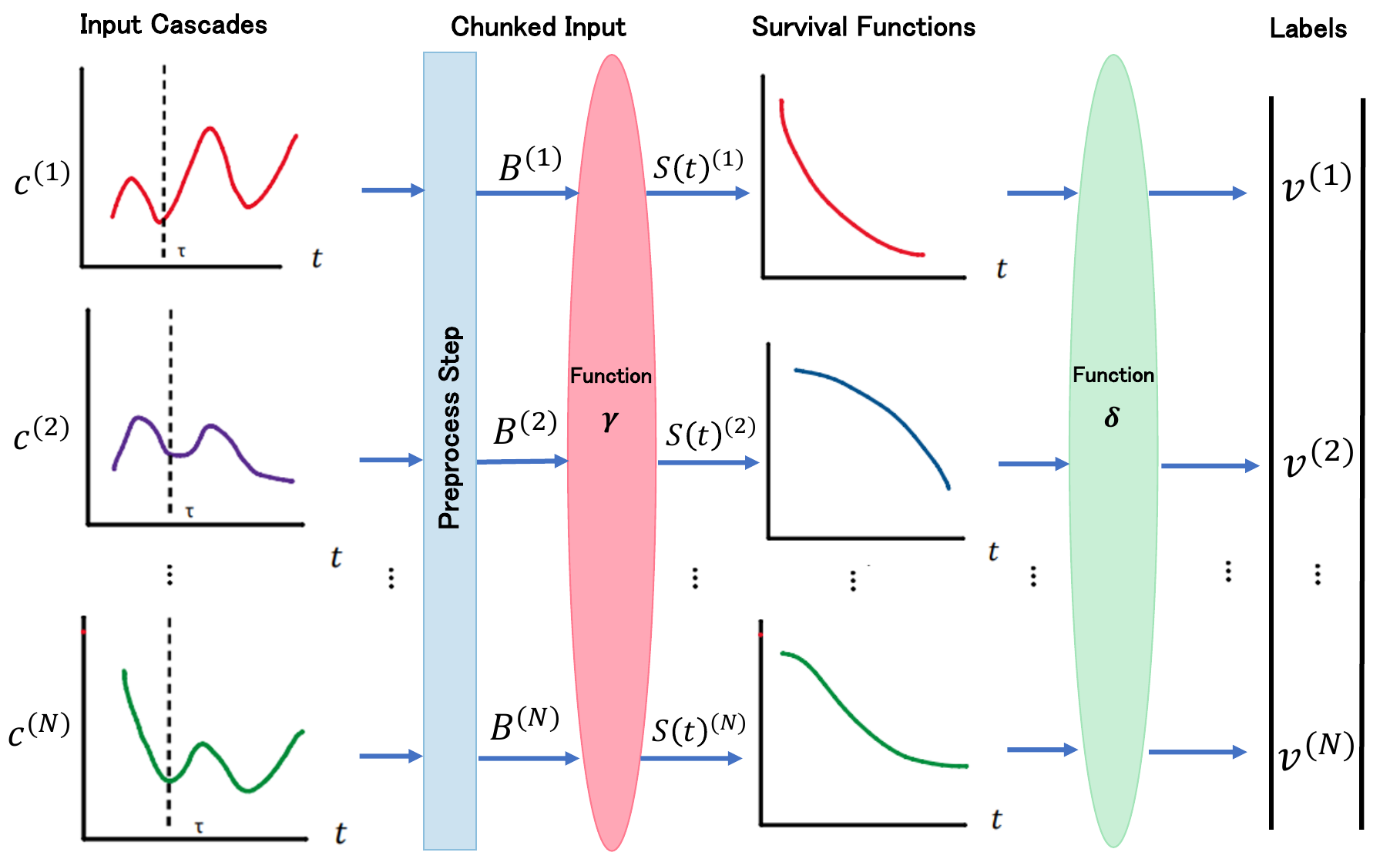}
\caption{A brief overview of the VEDSA method. A cascade of inputs is observed between $[0, \tau]$, corresponding to steps $0$ to $r$ after prepossessing by Figure \protect \ref{fig:input}. From the early stage of uncensored data, $\gamma$ fits the survival function and predicts the probability of virality for censored data. A discriminator $\delta$ will learn to output the virality label by classifying the cascades based on the estimated survival functions.}
\label{fig:overview}
\end{figure*}

\subsection{$\gamma$ Model: Survival Function Fitting}\label{subsec:ProblemDefinition}
To fit a good survival function, we have three choices \cite{xu2023coxnam,fitting2018}: 

\begin{enumerate}
\item 
Non-parametric: A function is learned without assumptions about distributions and without incorporating covariates, making it more complex as the number of observations increases. As a result, they have greater flexibility. A fitted non-parametric survival function may be a piecewise constant in the interval without observation. In this case, the survival function is called a non-smooth survival function, leading to less stable parameter estimation and making it more challenging to compare different survival functions. In this category, Kaplan-Meier \cite{kaplan1958nonparametric} is one of the best-known methods. 

\item
Semi-parametric: Cox's survival analysis \cite{cox1972regression} and its extension methods are based on this category. Unlike non-parametric methods, they make certain parametric assumptions, such as log-linear or linear relations between covariates and time to event. Additionally, Cox's regression estimates hazard ratios, providing insights into relative risks, but it does not directly predict absolute time to event. The estimated functions from this approach are non-smooth, like the non-parametric category. New attempts try to change the linear relation of covariates by adding nonentities to Cox using a mixture of experts \cite{nagpal2019nonlinear} or utilizing the neural network approaches such as MLP \cite{katzman2018deepsurv} and RNN \cite{giunchiglia2018rnn}.

\item 
Parametric: An approach to modeling time-to-event data that specifies a parametric form for the survival function. Parametric models assume that the survival distribution follows a specific functional form, unlike nonparametric methods. Hence, the parametric survival model is often used in situations where the shape of the survival curve is of interest and appropriate distributional assumptions are made. As a result of parametric models, survival curves can be extrapolated beyond the observed data, enabling us to predict future survival probabilities based on the observed data. Therefore, parametric models can handle censored data. The likelihood function is adjusted to account for censored observations in the estimation process. Even though these models are less flexible than non-parametric models, they are helpful in situations with limited training data. On the other hand, by choosing a continuous function, we can have a smooth survival function. Also, like in semi-parametrics, the incorporation of covariates can be used.
\end{enumerate}

The parametric approach is, therefore, the most effective method for detecting viral events based on survival function, as summarized in Table \protect \ref{tbl:fittingapproaches}. By observing the limited first part of the distribution, you can estimate the future of the survival function.  

\begin{table*}[ht!]
\centering % Center the table within the page
\caption{A summary of various approaches to fitting survival functions based on comparisons from different aspects. \ding{51} and \ding{55} refer to having or not having that feature. \ding{107} indicates that different methods in this category can take either a linear or non-linear approach. The \ding{115} demonstrates that this property does not generally exist but can be incorporated into deep-based methods if smooth activation functions are chosen. \ding{117} signifies smoothness by utilizing continuous distributions and smooth activation functions in deep methods.}
\label{tbl:fittingapproaches}
\resizebox{0.95\textwidth}{!}{
    \begin{tabular*}{\textwidth}{@{\extracolsep{\fill}} lccc} % Adjusted to use a width of 100% of the text width
    \toprule
    \textbf{Characteristics}        & \textbf{Non-parametric} & \textbf{Semi-parametric} & \textbf{Parametric}  \\
    \midrule
    Distribution Assumption         & \ding{55}      & \ding{55}       & \ding{51}   \\
    Incorporating of Covariates     & \ding{55}      & \ding{107}     & \ding{107}  \\
    Smoothness                      & \ding{55}      & \ding{115}     & \ding{117}  \\
    Predict Future                  & \ding{55}      & \ding{55}       & \ding{51}   \\
    Robustness to Data Limitations  & \ding{55}      & \ding{51}       & \ding{51}   \\ 
    \bottomrule
    \end{tabular*}
}
\end{table*}

If $t^{(i)}_{v}$ is the time of viral start for cascade $c^{(i)}$, we define $\sigma^{(i)}_{j}$ as an indicator that in each chunk does the cascade is in the viral state or not:

\begin{equation}
    \sigma^{(i)}_{j} =
    \begin{cases}
        1 & \text{cascade is in viral state at time $j \geq t^{(i)}_{v}$} \\
        0 & \text{\parbox[t]{.35\textwidth}{cascade does not reach viral state till time $j < t^{(i)}_{v}$ or the cascade is non-viral $n < \zeta$}}
    \end{cases}
\end{equation}

% \begin{equation}
%     \sigma^{(i)}_{j} =
%      \begin{cases}
% 1 &\text{cascade is in viral state at time $j \geq t^{(i)}_{v}$ }\\
% 0 &\text{cascade does not reach viral state till time $j < t^{(i)}_{v}$ or the cascade is non-viral $n < \zeta$ }
% \end{cases}
% \end{equation}

The objective function of these problems is to predict whether the number of participants in the social event cascade is less than $\zeta$ before time $j$ and becomes viral ($n \ge \zeta$) at time $j$ or whether it will occur later.

\begin{figure*}[ht]
\centering
\includegraphics[scale=0.55]{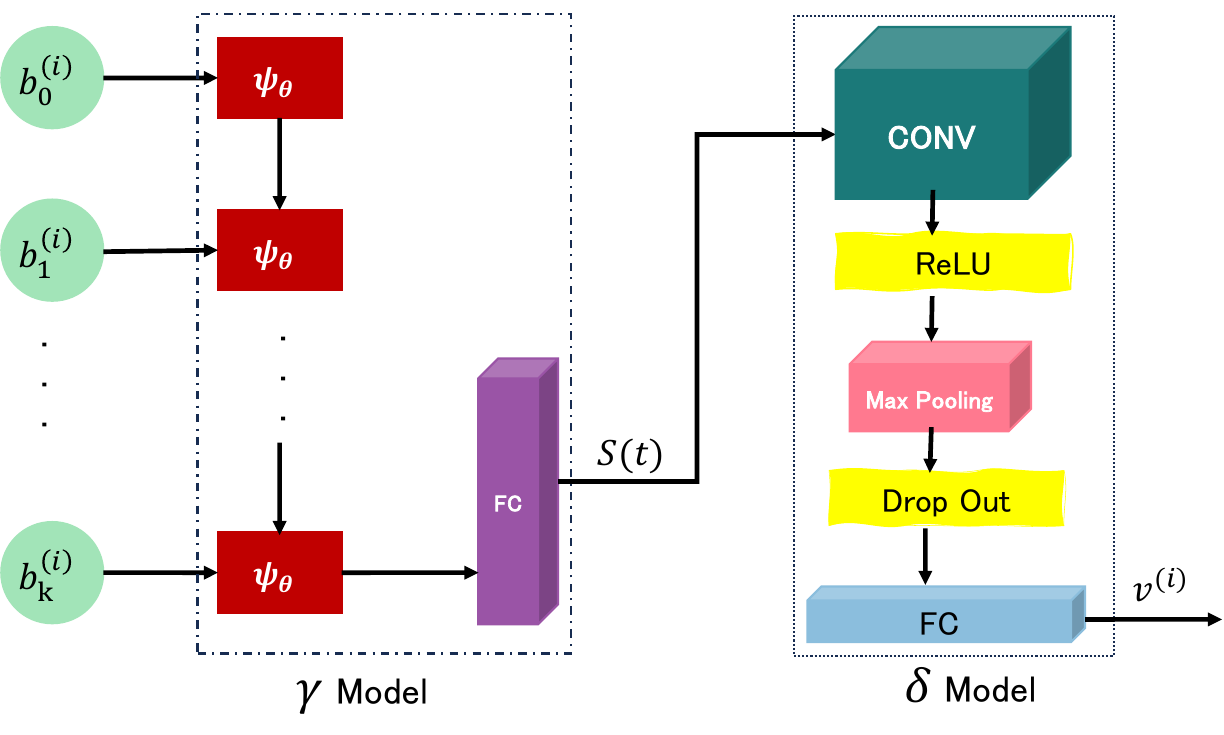}
\caption{VEDSA training phases. The first phase fits the survival function using our recurrent model, and the second phase uses the inferred survival function for viral detection. The red boxes ($\psi_\theta$) are cells of LSTMs.}
\label{fig:Architecture}
\end{figure*}

\begin{equation} \label{eq:mle_sample}
        \mathtt{L}^{(i)} =\prod_{0 \leq j \leq \frac{n}{\mathcal{L}}} \left(S^{(i)}(j)\right)^{1-\sigma^{(i)}_j} \left(f^{(i)}(j)\right)^{\sigma^{(i)}_{j}} 
\end{equation}

Considering $D$ different training independent and identically distributed (i.i.d.) cascades, the loss function is defined using the negative log-likelihood as follows:

\begin{align} 
\mathtt{L} &= -\ln{\left(\prod_{1 \leq i \leq D}\mathtt{L}^{(i)}\right)} \nonumber \\
&= \sum_{i=1}^{D} \sum_{j=1}^{\frac{n}{\mathcal{L}}} \left[ {(\sigma^{(i)}_j-1)} \ln{(S^{(i)}(j))} -\sigma^{(i)}_{j} \ln{(f^{(i)}(j))} \right] \label{eq:loss_total}
\end{align}

Utilizing Equation \ref{eq:survival_hazard_pdf}, the final loss function is:
 \begin{equation} \label{eq:loss_fitting_final}
	 \mathtt{L}=  \sum_{i=1}^{D} \sum_{j=1}^{\frac{n}{\mathcal{L}}} \left(
	 -{\sigma^{(i)}_j} 
	 \ln{\left(h^{(i)}(j)\right)} - \ln{\left(S^{(i)}(j)\right)} \right)
	 \end{equation}
	
Then, a parametric survival can be learned with the defined loss by choosing a suitable continuous distribution $f(t,\phi)$ where $\phi$ is the set of distribution unknown parameters. We try to learn these parameters based on observed input $B$ using a neural network whose parameters are $\theta$. Therefore, the distribution density is $f(t,\psi_{\theta}(B))$. Function $\psi$ is learning the non-linearity of covariates. For capturing the temporal correlation and sequential dependencies between these covariates $b^{(i)}_{j}$s, we utilized a Long Short-Term Memory network (LSTM). The last layer of LSTM is fed to a Fully Connected (FC) network layer. After applying exponential or soft-plus activation functions to the output of the FC layer, smooth $S(t)$ is obtained. According to data-driven statistical analysis of real-world data, \cite{yu2015micro} demonstrates that cascade behaviors in social media are governed by three well-known Exponential, Rayleigh, and Weibull distributions.

In conclusion, the $\gamma$ model is an LSTM network with a fully connected layer trained with loss function Equation \ref{eq:loss_fitting_final}. The hazard and survival functions for the relative parametric continuous distribution are given in Table \ref{tbl:distributions}. Based on different distribution assumptions, three versions of the proposed method will be used in the rest of the paper named: VEDSA-Exponential, VEDA-Rayleigh, and VEDSA-Weibull.

\begin{table*}[h!]
\centering
\caption{A summary of various distributions and their properties.}
\label{tbl:distributions}
\begin{tabular}{@{} CCCCC @{}} % Use the newly defined centered column type
\toprule
\textbf{Distribution} & \textbf{PDF} & \textbf{CDF} & \textbf{Survival Function ($S(T)$)} & \textbf{Hazard Function ($h(t)$)} \\
\midrule
Exponential & $\lambda \exp(-\lambda t)$ & $1-e^{-\lambda t}$ & $e^{-\lambda t}$ & $\lambda$ \\
Rayleigh & $\alpha t \exp(-0.5 \alpha t^2)$ & $1-\exp(-0.5 \alpha t^2)$ & $\exp(-0.5 \alpha t^2)$ & $\alpha t$ \\
Weibull & $\frac{\kappa}{\lambda}\exp\left(-\left( \frac{t}{\lambda} \right)^{\kappa}\right)\left(\frac{t}{\lambda}\right)^{\kappa-1}$ & $1-\exp\left(-\left( \frac{t}{\lambda} \right)^{\kappa}\right)$ & $\exp\left(-\left( \frac{t}{\lambda} \right)^{\kappa}\right)$ & $\frac{\kappa}{\lambda}\left(\frac{t}{\lambda}\right)^{\kappa-1}$ \\
\bottomrule
\end{tabular}
\end{table*}

\subsection{$\delta$ Model: Discriminator}\label{subsec:Discriminator}

The predictor model aims to classify the cascade as viral or non-viral. It achieves this by using the inferred survival function derived from the initial model. The inferred survival function is learned through the use of convolutional and pooling layers. ReLU is used to apply a non-linear activation function. In order to prevent overfitting, dropout is employed as a regularization technique. Finally, once the features have been flattened into a one-dimensional vector, they will be forwarded through fully connected layers to make the final prediction $v^{(i)}$ as illustrated in Fig.  \ref{fig:Architecture}.

\section{Numerical Results}
We first describe the real datasets, along with a statistical analysis to highlight their characteristics. After that, we will evaluate the performance of VEDSA using various distributions under different conditions. A challenging aspect of viral event detection is the censored data, which significantly impacts the learning process and the accuracy of future predictions. The in-depth examination provided in different censcored data situations will contribute to our understanding of the durability and flexibility of our proposed model. Moreover, we will discuss the competitor baselines and thoroughly explain their choice. We will then conduct a comparative analysis of the results obtained from our proposed method against these baselines, providing insights into the effectiveness and advantages of our approach.
\subsection{Datasets}
We used different social networks, including two microblogging  and one social news aggregator. The information spread mechanism over each of them is different. We categorized the number of events from the beginning to the end of cascade age $n$ based on predefined thresholds named $\zeta_1$ and $\zeta_2$ into three distinct types for virality definition: when $n$ falls within a range from 0 to a specified value $\zeta_1$, shows non-viral cascades. Conversely, cascades categorized as viral occur when the engagement exceeds a defined minimum threshold $\zeta_2$. Our study does not focus on intermediate cascades, which lie within the range of $\zeta_1<n<\zeta_2$.

\textbf{Twitter Dataset}
We utilized a dataset of 166,076 tweets collected from Twitter during the period between October 7 and November 7, 2011. This dataset was originally made available by a previous study (Zhao et al. 2015) through the SNAP project at Stanford University (http://snap.stanford.edu/seismic/). The dataset contains key details for each tweet, including the tweet ID, the time it was posted, the times at which retweets occurred, and the number of followers associated with both the original tweet and its retweets. Retweet timings were tracked up to 7 days (168 hours) after the original post. Although the dataset provides basic network information, such as follower counts (accessible through the Twitter API), it does not include the full structure of user connections.

\textbf{Weibo Dataset}
The Weibo dataset comprises user-generated content from Sina Weibo, one of the leading social media services in China. This dataset (https://www.aminer.cn/influencelocality), spanning from August 30, 2009, to October 26, 2012, includes interactions from approximately 1.78 million users, generating around 23 million posts. Notably, within this dataset, there are approximately 300,000 original posts that have evolved into information cascades. We carefully sampled these cascades for our study, as detailed in Table \ref{tbl:DatasetDescp}, to analyze the patterns of information spread across the platform.

\textbf{Digg Dataset}
The Digg dataset is sourced from the Digg platform, a well-known social news aggregator where users can submit articles and vote on them. This dataset (https://www.isi.edu/people-lerman/research/downloads/) includes a collection of user interactions over a specified period, reflecting voting records for 3553 stories promoted to the front page over a period of a month in 2009. The voting record for each story contains id of the voter and time stamp of the vote. In addition, data about friendship links of voters was collected. 

\begin{table*}[ht]
\centering
\caption{Dataset Information}
\label{tbl:DatasetDescp}
% Use tabular instead of tabular* if you don't need to stretch the table
\begin{tabular}{lllll} % Adjust column specifiers as needed (e.g., l for left-aligned)
\toprule
\textbf{Dataset} & \textbf{Range}      & \textbf{Year} & \textbf{\#Cascades} & \textbf{\#Reshares}  \\ 
\midrule
Twitter & Oct.7th - Nov.7th & 2011 & 166076 & 34784488 \\
Digg    & June & 2009 & 3553 & 3018197 \\
Weibo & Aug.30th - Dec.26th & 2009 - 2012 & 30000 & 23755810 \\
\bottomrule
\end{tabular}
\end{table*}

\subsection{Effect of Distribution}
In Twitter,  the model demonstrates the highest accuracy and F1 scores with the Weibull distribution, achieving an accuracy of $92.51\%$ and an F1 score of $92.23\%$ within the first $2$ hours. Performance remains consistently strong across the subsequent time intervals, peaking at $96.55\%$ accuracy and $96.56\%$ F1 score at the $24$ hour mark with the Weibull distribution.
In Digg, the results indicate a notable difference in performance compared to the Twitter dataset, with the Weibull distribution again leading at the 2-hour mark with $59.51\%$ accuracy and $52.83\%$ F1 score.
The overall performance is lower than that of Twitter, highlighting the varying dynamics of information spread within different social media contexts. The best accuracy recorded is $97.37\%$ at 24 hours with the Weibull distribution.
For Weibo, the model achieves a maximum accuracy of $83.56\%$ with the Weibull distribution at the 10-hour interval, indicating effective engagement with the platform's user interactions.
The results for Weibo are relatively stronger than those for Digg but not as high as Twitter, emphasizing the unique characteristics of user behavior and content sharing on this platform.
Overall, the findings underscore the influence of the chosen distribution on the model's performance across different datasets. The Weibull distribution consistently outperforms the others in terms of accuracy and F1 scores, suggesting that it is the most suitable for modeling the dynamics of information spread in these social media contexts. These insights are valuable for refining our approach to understanding how information cascades develop over time.

\begin{table*}[ht]
\centering
\caption{Model Metrics for Non-Viral and Viral Cascades}
\label{tab:model_metrics}
\begin{tabular}{@{} lllllllll @{}} % Adjusted to standard column types for simplicity
\toprule
\multirow{2}{*}{\textbf{Dataset}} & \multirow{2}{*}{\textbf{Model Name}} & \multicolumn{2}{c}{\textbf{Precision}} & \multicolumn{2}{c}{\textbf{Recall}} & \multicolumn{2}{c}{\textbf{F1}} & \multirow{2}{*}{\textbf{Accuracy}} \\
                                  &                                     & \textbf{Non-Viral} & \textbf{Viral} & \textbf{Non-Viral} & \textbf{Viral} & \textbf{Non-Viral} & \textbf{Viral} &                                 \\ 
\midrule
\multirow{4}{*}{Twitter (8.33\%)} & Seismic                     & 72.31            & $\mathbf{100}$ & $\mathbf{100}$ & 61.70            & 83.93            & 76.32            & 80.85                      \\
                                  & Cheng                       & 93.10            & 95.60          & 95.85          & 92.69            & 94.46            & 94.12            & 94.30                      \\
                                  & EDRN                        & 91.92            & 96.11          & 96.34          & 91.44            & 94.08            & 93.71            & 93.90                      \\
                                  & Linear                      & 77.08            & 96.37          &97.37          &70.72              & 86.05            & 81.58            & 84.12
                                    \\
                                  & VEDSA           & $\mathbf{94.61}$            & 96.34          & 96.45          &$\mathbf{94.44}$         & $\mathbf{95.52}$ & $\mathbf{95.38}$ & $\mathbf{95.45}$           \\
                                  % & VEDSA-Exponential        & $\mathbf{95.38}$ & 94.50          & 94.51          & $\mathbf{95.37}$ & 94.94            & 94.39            & 94.94                      \\
                                  % & VEDSA-Reyleigh           & 94.48            & 93.90          & 93.94          & 94.44            & 94.21            & 94.17            & 94.19                      \\ 
\midrule
\multirow{4}{*}{Digg (1.94\%)}    & Seismic                     & 57.16            & $\mathbf{100}$ & $\mathbf{100}$ & 25.07            & 72.75            & 40.08            & 62.53                      \\
                                  & Cheng                       & 76.43            & 80.27          & 80.53          & 76.12            & 78.43            & 78.14            & 78.28                      \\
                                  & EDRN                        & 71.84            & 93.88          & 96.10          & 61.33            & 82.22            & 74.19            & 78.95                      \\
                                  & Linear                      & 66.50            & 78.50          & 85.06          & 56.00            & 74.64            & 65.37            &70.72
                                  \\
                                  & VEDSA            & $\mathbf{85.63}$ & 91.97          & 92.86          & $\mathbf{84.00}$            & $\mathbf{89.10}$ & $\mathbf{87.80}$ & $\mathbf{88.49}$           \\
                                  % & VEDSA-Exponential        & 69.95            & 78.51          & 83.12          & 63.33            & 83.12            & 70.11            & 73.36                      \\
                                  % & VEDSA-Reyleigh           & 83.80            & 78.40          & 77.27          & $\mathbf{84.67}$ & 80.41            & 81.41            & 80.92                      \\

\midrule
\multirow{4}{*}{Weibo (0.972\%)}    & Seismic                     & 65.67            & $\mathbf{99.89}$ & $\mathbf{99.95}$ & 47.75   &79.26            & 64.61            & 73.85                     \\
                                  & Cheng                       & 84.92            & 94.75          & 95.5          & 82.72            & $\mathbf{89.9}$            & 88.32           & 89.17                      \\
                                  & EDRN                        & 85.6           & 93.35         & 94.13          & 83.88            & 89.66           & 88.36            & 89.05                     \\
                                  & Linear                      & 77.98           & 96.83         & 97.69         & 71.92            & 86.73           & 82.54           & 84.92
                                  \\
                                  & VEDSA          & $\mathbf{86.77}$    & 92.38          & 93.08          & $\mathbf{85.54}$   & 89.82  & $\mathbf{88.83}$ & $\mathbf{89.34}$           \\
                                  % & VEDSA-Exponential        & 69.95            & 78.51          & 83.12          & 63.33            & 83.12            & 70.11            & 73.36                      \\
                                  % & VEDSA-Reyleigh           & 83.80            & 78.40          & 77.27          & $\mathbf{84.67}$ & 80.41            & 81.41            & 80.92                      \\
                                  
\bottomrule
\end{tabular}
\end{table*}

\begin{table*}[ht]
\centering
\caption{Comparison of VEDSA on different distributions}
\label{tab:model_metrics}
\makebox[\textwidth]{ % Centering the table explicitly
    \resizebox{1.35\textwidth}{!}{%
    \begin{tabularx}{1.35\textwidth}{l l *{12}{>{\centering\arraybackslash}X}} 
    \toprule
    \multirow{2}{*}{\textbf{Dataset}} & \multirow{2}{*}{\textbf{Distribution}} & \multicolumn{2}{c}{\textbf{2 hours}} & \multicolumn{2}{c}{\textbf{6 hours}} & \multicolumn{2}{c}{\textbf{10 hours}} & \multicolumn{2}{c}{\textbf{14 hours}} & \multicolumn{2}{c}{\textbf{18 hours}} & \multicolumn{2}{c}{\textbf{24 hours}} \\
                                      &                                        & \textbf{Accuracy} & \textbf{F1} & \textbf{Accuracy} & \textbf{F1} & \textbf{Accuracy} & \textbf{F1} & \textbf{Accuracy} & \textbf{F1} & \textbf{Accuracy} & \textbf{F1} & \textbf{Accuracy} & \textbf{F1} \\
    \midrule
    \multirow{3}{*}{Twitter} 
    & Weibull     & \textbf{92.51}  & \textbf{92.23}  & 93.10 & 92.96 & 93.38 & 93.74 & 95.45 & \textbf{95.38}  & \textbf{95.80} & \textbf{95.79}   & 96.55 & 96.56 \\
    & Exponential & 90.62  & 90.43  & \textbf{93.67} & \textbf{93.59} & \textbf{94.30} & \textbf{94.25} & \textbf{94.53} & 94.51  & 95.40 & 95.38   & \textbf{96.72} & \textbf{96.72} \\
    & Rayleigh    & 49.53  & 66.25  & 49.71 & 66.41 & 49.71 & 66.41 & 49.71 & 66.41  & 49.71 & 66.41   & 49.71 & 66.41 \\
    \midrule
    \multirow{3}{*}{Digg} 
    & Weibull     & \textbf{59.51}  & \textbf{52.83}  & \textbf{70.07} & \textbf{67.62} & \textbf{79.28} & \textbf{78.05} & \textbf{88.49} & \textbf{87.80}  & \textbf{92.11} & \textbf{91.67}   & \textbf{97.37} & \textbf{97.28} \\
    & Exponential & 52.96  & 12.27  & 65.79 & 49.02 & 76.32 & 70.25 & 83.88 & 81.23  & 89.47 & 88.49   & \textit{86.18} & \textit{86.00} \\
    & Rayleigh    & 49.34  & 66.08  & 49.34 & 66.08 & 49.34 & 66.08 & 49.34 & 66.08  & 49.34 & 66.08   & 49.34 & 66.08 \\
    \midrule
    \multirow{3}{*}{Weibo} 
    & Weibull     & \textbf{79.04}  & \textbf{77.14}  & 84.37 & 83.56 & 88.03 & \textbf{87.47} & 89.34 & 88.83  & 91.27 & 90.81   & 93.18 & 92.78 \\
    & Exponential & 78.95  & 76.30  & \textbf{85.33} & \textbf{83.78} & 88.03 & 87.40 & \textbf{89.87} & \textbf{89.38} & \textbf{91.97} & \textbf{91.64}   & \textbf{93.36} & \textbf{93.00} \\
    & Rayleigh    & 49.34  & 66.08  & 49.34 & 66.08 & 49.34 & 66.08 & 49.34 & 66.08  & 49.34 & 66.08   & 49.34 & 66.08 \\
    \bottomrule
    \end{tabularx}%
    }
}
\end{table*}

\section{Conclusion}
In this paper, we aimed to interpret the issue of partial observations in dissemination as the availability of an initial segment of the dissemination data stream. To this end, we proposed a deep network-based approach for learning a parametric survival function. Through this process, we first accounted for the non-linearity of the correlated input variables. Second, we were able to consider the temporal dependencies of the sequential input data. Third, due to the chosen parametric distribution, we could learn the survival function even with limited initial data and utilize it for future predictions. Additionally, we demonstrated that the obtained survival functions could be classified for detecting viral and non-viral cascades.

\newpage

\bibliographystyle{natbib}
\bibliography{refs}

\appendix

\end{document}